\documentclass[%
 reprint,
 superscriptaddress,
 longbibliography,
 amsmath,amssymb,
 aps,
 prd,
]{revtex4-2}

\usepackage{graphicx}
\usepackage{dcolumn}
\usepackage{bm}
\usepackage{xcolor}
\usepackage{orcidlink}

\DeclareRobustCommand{\VAN}[3]{#2}
\let\VANthebibliography\thebibliography
\def\thebibliography{\DeclareRobustCommand{\VAN}[3]{##3}\VANthebibliography}

\def\aap{A\&A}
\def\apj{ApJ}

\def\apjl{ApJ}
\def\mnras{MNRAS}
\def\araa{ARA\&A}

\def\apjs{ApJS}
\def\prd{Phys. Rev. D}
\def\pasp{PASP}
\def\zap{ZAP}

\begin{document}

\preprint{APS/123-QED}

\title{The Impact of Neutrino Magnetic Moments on the Evolution of the Helium Flash and Lithium-Rich Red Clump Stars}

\author{Xizhen Lu}
\affiliation{School of Physical Science and Technology, Xinjiang University, Urumqi, 830046, China} 
\author{Chunhua Zhu}
\affiliation{School of Physical Science and Technology, Xinjiang University, Urumqi, 830046, China}

\author{Guoliang L\"{u}}
\email{guolianglv@xao.ac.cn}
\affiliation{School of Physical Science and Technology, Xinjiang University, Urumqi, 830046, China}
\affiliation{Xinjiang Astronomical Observatory, Chinese Academy of Sciences, 150 Science 1-Street, Urumqi, Xinjiang 830011, China}

\author{Sufen Guo}
\affiliation{School of Physical Science and Technology, Xinjiang University, Urumqi, 830046, China}

\author{Zhuowen Li}
\affiliation{School of Physical Science and Technology, Xinjiang University, Urumqi, 830046, China}

\author{Gang Zhao}
\affiliation{CAS Key Laboratory of Optical Astronomy, National Astronomical Observatories, Beijing 100101, China}

\date{\today}

\begin{abstract}
The detection of the neutrino magnetic moment (NMM,$\mu_v$) is one of the most significant challenges in physics. 
The additional energy loss due to NMM can significantly influence the helium flash evolution in low-mass stars. Therefore, studying this process serves as the most robust method to constrain NMM.
Using the Modules for Experiments in Stellar Astrophysics (MESA) stellar evolution code, we investigated the impact of NMM on the helium flash evolution in low-mass stars. We found that NMM leads to an increase in both the critical helium core mass required for the helium flash and the luminosity of the tip of the red giant branch (TRGB). For a typical $Z = 1 Z_{\odot}$ , $M$ = 1.0 $M_{\odot}$, and $\mu_v = 3 \times 10^{-12} \mu_{\mathrm{B}}$ model, the helium core mass increases by $\sim 5\%$, and the TRGB luminosity increases by $\sim 35\%$ compared to the model without NMM. 
However, contrary to previous conclusions, our model indicates that the helium flash occurs earlier, rather than delayed, with increasing NMM values.
This is because the additional energy loss from NMM accelerates the contraction of the helium core, releases more gravitational energy that heats the hydrogen shell and increases the hydrogen burning rate, thereby causing the helium core to reach the critical mass faster and advancing the helium flash.
An increase in NMM results in a higher peak luminosity for the first helium flash, a more off-center ignition position, and sub-flashes with higher luminosities, shorter intervals, and higher frequency.
We found that the internal gravity wave (IGW) mixing generated by the helium flash can induce sufficient mixing in the radiative zone, turning the overshoot region into a low-Dmix bottleneck within the stellar interior. The increase in NMM in the model narrows the overshoot bottleneck region, enabling Li to enter the surface convection zone more quickly, thereby enhancing the enrichment effect of IGW mixing on surface Li. For models incorporating both NMM and IGW mixing, the reduction in the overshoot bottleneck region allows them to effectively produce super Li-rich red clump star samples.
    
\end{abstract}

\maketitle


\section{Introduction}
In the Standard Model, neutrinos are described as electrically neutral leptons without rest mass. However, neutrino oscillations, which demonstrate that neutrinos have rest mass, suggest that neutrinos possess properties beyond the Standard Model. Some extensions of the Standard Model provide varying predictions for non-zero mass neutrinos possessing a neutrino magnetic moment (NMM)\cite{Fujikawa1980PhRvL,Dvornikov2004PhRvD}. Majorana neutrinos, in particular, have a higher upper limit for NMM compared to Dirac neutrinos, indicating that measuring NMM is a theoretical and experimental pathway to uncovering exotic physics\cite{Barbieri1988NuPhB,Bell2005PhRvL,Davidson2005PhLB,Bell2006PhLB}.

Currently, there are two approaches to constraining the NMM. One approach is experimental, involving the measurement of the electron recoil spectrum in (anti) neutrino-electron scattering\cite{Reines1976PhRvL,Li2003PhRvL,Daraktchieva2005PhLB,Deniz2010PhRvD,Agostini2017PhRvD}, with the latest result from XENONnT yielding $\mu_v<6.5 \times 10^{-12} \mu_{\mathrm{B}}$ ($\mu_B=e / 2 m_e$ denotes the Bohr magneton.)\cite{Aprile2022PhRvL}. The other approach relies on astronomical observations to provide constraints on the upper limit of NMM. A non-zero NMM can influence various neutrino cooling mechanisms (pair, plasma, photon, bremsstrahlung), particularly as NMM couples neutrinos to photons through an effective Lagrangian term, resulting in significant changes in the emission rate of plasma processes($\gamma \rightarrow \bar{\nu} \nu$)\cite{Haft1994ApJ,Heger2009ApJ,Mori2021}. 
The vast size of stars amplifies the small energy loss differences from individual reactions, allowing them to significantly impact macroscopic properties.  These effects are particularly detectable in processes like the helium flash, where neutrino-driven energy loss dominates.
Several previous studies have proposed that a non-zero NMM would delay the helium flash\cite{Arceo2015APh,Giunti2016AnP}. This has led to constraints on the luminosity of low-mass stars just before the helium flash, the tip of the red giant branch (TRGB), providing more stringent constraints than experimental measurements.\cite{Arceo2015APh,Serenelli2017AA}. Currently, the best constraints on the luminosity of the TRGB in globular clusters are $\mu_v<1.5 \times 10^{-12} \mu_{\mathrm{B}}$\cite{Capozzi2020PhRvD}. With improvements in telescope precision and reductions in observational errors, stellar evolution theory is now starting to dominate the overall uncertainty, limiting further improvements in constraint precision\cite{Capozzi2020PhRvD}.

The helium flash is a brief but intense helium fusion event that occurs in the cores of stars with masses ranging from $0.7\lesssim M/M_\odot\lesssim 2$ as they evolve to the TRGB. When low-mass stars evolve into the red giant phase, they experience a contraction of their helium cores under the dominance of self-gravity until electron degeneracy pressure is sufficient to maintain equilibrium\cite{Sweigart1978}. As hydrogen burning continues in the shell surrounding the helium core, the mass and temperature of the helium core increase. Due to the isothermal nature of degenerate matter, the entire helium core can be regarded as an isothermal core. When the temperature of the helium core reaches the critical ignition temperature, the entire core undergoes an explosive combustion, known as the helium flash\cite{Schwarzschild1962}. Theoretical predictions indicate that the central region of the helium core, where density is highest, will cool due to the energy carried away by copious neutrino emission, causing helium ignition to occur in regions away from the center. The immense energy released in the ignition zones relieves the degeneracy of these regions and causes them to expand outward, at which point the temperature gradient surpasses the adiabatic gradient, triggering convection. At the end of the first helium flash, the region below the helium ignition zone remains in an unignited degenerate state. This leads to several subsequent weaker sub-flashes, each occurring progressively closer to the center, lifting the degeneracy and igniting the helium in those regions. This process continues until the final sub-flash reaches the center, fully lifting the degeneracy of the entire helium core and allowing it to burn stably.

Clearly, the physical process that most directly governs the evolution of the helium flash is the energy loss mechanism mediated by neutrinos. The neutrinos generated at the center of the degenerate helium core are all thermal process neutrinos, predominantly plasma decay neutrinos. Therefore, the additional energy loss due to a non-zero NMM inevitably results in a larger critical degenerate helium core mass\cite{Arceo2015APh,Serenelli2017AA}. Currently, all constraints on NMM derived from TRGB observations are obtained using the helium core mass-luminosity relationship provided by stellar evolution codes\cite{Arceo2015APh,Capozzi2020PhRvD}.

Observations of lithium-rich giants pose a challenge to stellar evolution theory\cite{Brown1989ApJS,Kumar2011ApJ,Yan2018NatAs} . During the red giant phase, the inward extension of the convective envelope leads to significant lithium depletion. As a result, standard theoretical models predict that low-mass red giants should exhibit very low surface lithium abundances, with a logarithmic lithium abundance index $A(\mathrm{Li})$ typically below 1.5 $\mathrm{dex}$\footnote{Here, $A(\mathrm{Li}) = \log _{10}(n(\mathrm{Li}) / n(\mathrm{H}))+12$, where $n(\mathrm{Li})$ and $n(\mathrm{H})$ are the number densities of Li and H, respectively.} Therefore, giants with 
$A(\mathrm{Li})>1.5 \mathrm{dex}$ are classified as Li-rich giants, while those with 
$A(\mathrm{Li})>3.2 \mathrm{dex}$ are referred to as super Li-rich giants.
Recently, observations have revealed that the majority of Li-rich and super Li-rich giants are red clump stars that have recently undergone a helium flash\cite{Kumar2020NatAs,Yan2021NatAs}. Schwab \cite{Schwab2020} proposed a formation mechanism for Li-rich red clump stars (RC), wherein the helium flash triggers intense mixing by internal gravity waves(IGW). 3D simulations indicate that the intensity of IGW is primarily proportional to the helium luminosity during the helium flash. The magnitude of the NMM can significantly affect the process of energy loss, thereby influencing the helium luminosity during the helium flash. This means that we can constrain the numerical value of the NMM by comparing observational samples of Li-rich red clump stars with models.

In this paper, we plan to investigates the influence of the NMM on the helium flash evolution in low-mass stars, with a focus on the effects of NMM on the critical helium core mass, the onset time of the helium flash, and the TRGB luminosity. We also examine the role of IGW mixing triggered by the helium flash in enhancing the surface lithium abundance, particularly in NMM-involved evolution, and make a comparison with a sample of lithium-rich red clump stars.
Section 2 provides a detailed introduction to the software tools, input physical parameters, reaction networks, and model selection. Section 3 discusses the changes in evolutionary outcomes under different solar masses, metallicities, and neutrino magnetic moments, along with the impact of different mixing parameters on the evolution of stellar surface elemental abnormalities. Section 4 presents the conclusions.

\section{MODEL}
We employ the open-source stellar evolution code (\texttt{MESA}, version 15140, \cite{paxton2011, paxton2013, paxton2015, paxton2018, paxton2019}) to construct 1 to 2 $\rm M_{\odot}$ stellar models.
The mass loss formula in \cite{Reimers1975} is adopted, with the loss coefficient ($\eta$) set to 0.3. Convection boundaries are calculated using the Ledoux criterion, and a mixing-length parameter of 1.8 is adopted \cite{Lv2006, Gao2022A&A}. The overshoot area is calculated by the step-function overshooting which depends on the overshooting parameter $\alpha_{\rm ov}$. Following \cite{Gao2022A&A}, $\alpha_{\rm ov}=0.335$ in our simulations.
we also use the `element diffusion' option, which accounts for transport processes driven by pressure gradients (gravity), temperature gradients, composition gradients, and radiation pressure. This treatment primarily causes gravity sedimentation of heavy elements in the surface radiative zone\cite{paxton2018, paxton2019}. 
We adopt nuclear reaction rates compiled by \cite{Cyburt2010ApJS}. We use $\mathrm{MESA28.net}$ nuclear network which includes nuclear reactions from ${ }^{1} \mathrm{H}$ to${ }^{22} \mathrm{Ne}$. 
We simulate models with two metallicities: $Z = 1 Z_{\odot}$ and $Z = 0.01 Z_{\odot}$, where $Z_{\odot} = 0.014$\cite{Asplund2009ARAA}. 
For the initial $A(\mathrm{Li})=\log (\mathrm{Li} / \mathrm{H})+12$ is set to 2.8, which is identical to the A(Li) value found in meteorites\cite{Kumar2020NatAs,Schwab2020}. 

\subsection{Neutrino Magnetic Moment}
If neutrinos possess a non-zero NMM, the NMM couples neutrinos to photons via an effective Lagrangian term:
\begin{align}
	L=-\frac{1}{2} \mu^{i j} \bar{\psi}_i \sigma_{\alpha \beta} \psi_j F^{\alpha \beta}
	\label{eq:lagrangian}
\end{align}
where $\psi$ is the neutrino field, $F$ the electromagnetic field tensor, $\alpha$,$\beta$ are Lorentz indices, and $i$, $j$ are the flavor indices. Consequently, the mechanism of neutrino thermal production will acquire additional electromagnetic contributions from the interaction term in \ref{eq:lagrangian}. For low-mass stellar evolution, the primary processes involved are plasma and pair processes\cite{Mori2021}.
The energy loss rate due to the plasmon decay is given by
\begin{align}
	&&& \epsilon_{\text {plas }}^\mu=0.318\left(\frac{\omega_{\mathrm{pl}}}{10 \mathrm{keV}}\right)^{-2}\left(\frac{\mu_{12}}{10^{-12} \mu_{\mathrm{B}}}\right)^2 \epsilon_{\mathrm{plas}}
\end{align}
where $\rm \epsilon_{\text {plas }}$ is the standard plasmon decay rate \cite{Itoh1996ApJS} and  $\rm \omega_{\text {pl }}$ is the plasma frequency \cite{Raffelt1996}
\begin{align}
	&&& \omega_{\mathrm{pl}}=28.7 \mathrm{eV} \frac{\left(Y_{\mathrm{e}} \rho\right)^{\frac{1}{2}}}{\left(1+\left(1.019 \times 10^{-6} Y_{\mathrm{e}} \rho\right)^{\frac{2}{3}}\right)^{\frac{1}{4}}}
\end{align}
where $Y_{\mathrm{e}}$ represents the electron mole fraction, $\rm \rho$ denotes the density in units of $\rm g\ cm^{-3}$. The energy loss rate due to the pair production is given by \cite{Heger2009ApJ}
\begin{align}
	&&& \epsilon_{\text {pair }}^\mu=1.6 \times 10^{11} \mathrm{erg}^{-1} \mathrm{~s}^{-1}\left(\frac{\mu_{12}}{10^{-10} \mu_{\mathrm{B}}}\right)^2 \frac{e^{-\frac{118.5}{T_8}}}{\rho_4}
\end{align}
where $\rm T_{8}$ is the temperature in units of $\rm 10^{8}\ K$ and $\rm \rho_{4} =\rho/(10^{4} g\ cm^{-3})$. We added these two additional energy loss calculation formulas to $\rm run\_star\_extras$.

\subsection{Mixing by IGW}
IGW are believed to be excited by turbulent convective motions within stars\cite{Press1981ApJ, Lecoanet2013MNRAS}.The energy propagated outward by IGW disrupts the hydrostatic equilibrium of the radiative zone and leads to mixing in that region. The IGW model has successfully explained the variations in Li abundance observed in the Sun and low-mass stars\cite{Garcia1991ApJ, Charbonnel2005Sci}.
In the research results of \cite{Schwab2020} and \cite{Gao2022A&A}, it is hypothesized that IGW excited during the first helium flash in red giants can induce mixing in the radiative zone. They simulated the mixing caused by IGW by adding a fixed mixing coefficient and found that this mixing can effectively transport ${ }^{7} \mathrm{Be}$ from the hydrogen-burning region to the overlying convective zone. The ${ }^{7} \mathrm{Be}$ is then brought to the stellar surface and decays into ${ }^{7} \mathrm{Li}$, thus contributing to the explanation of lithium-rich red clump stars. However, simulating the mixing induced by IGW using a fixed mixing coefficient is highly inappropriate. Recently, \cite{Herwig2023MNRAS} obtained through 3D simulations that the mixing dmix induced by IGW is proportional to:
\begin{align}
	\mathrm{D}_{\mathrm{IGW}} \approx \eta \cdot \frac{(\nabla \times \mathbf{u})^2 \cdot K}{N^2}
\end{align}
The thermal diffusivity K is given by
\begin{align}
	K = \frac{4 a c T^3}{3 \kappa \rho^2 C_P}
\end{align}
and
\begin{align}
	\frac{(\nabla \times \mathbf{u})^2}{N^2} \propto L^{4 / 3}
\end{align}
Here, $N$ is again the \text { Brunt-Väisälä }(buoyancy) frequency, $a$ is the radiation constant, $c$ is the speed of light, $\kappa$ is the Rosseland mean opacity and $C_P$ is the specific heat at constant pressure, and the luminosity $L$ is the helium luminosity $L_{\mathrm{He}}$ at the helium core boundary. The dimensionless coefficient $\eta$ is set to 0.1 in this study, with its value determined by 3D hydrodynamical simulations\cite{Garaud2016ApJ,Prat2016A&A,Blouin2023MNRAS}. Therefore, following the approach of \cite{Denissenkov2024MNRAS}, we simplify the formula to :
\begin{align}
	\mathrm{D}_{\mathrm{IGW}} = K \left( \frac{L_{\mathrm{He}}}{10^{3} L_{\odot}} \right)^{4 / 3}
\end{align}

\begin{figure*}[htbp]
	\centering
	\begin{minipage}[b]{0.49\linewidth}
		\centering
		\includegraphics[width=\linewidth]{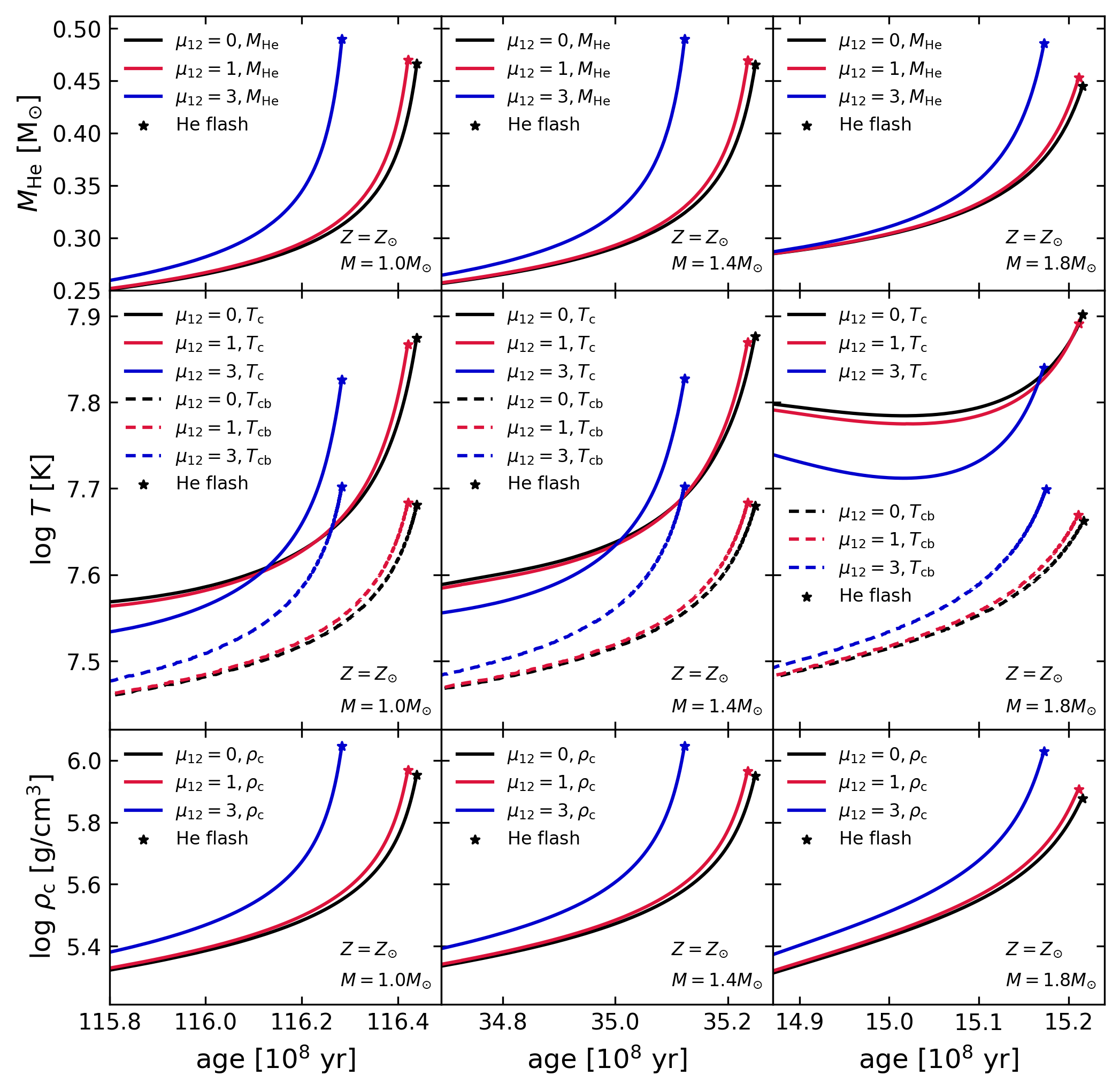}
	\end{minipage}
	\hfill	
	\begin{minipage}[b]{0.49\linewidth}
		\centering
		\includegraphics[width=\linewidth]{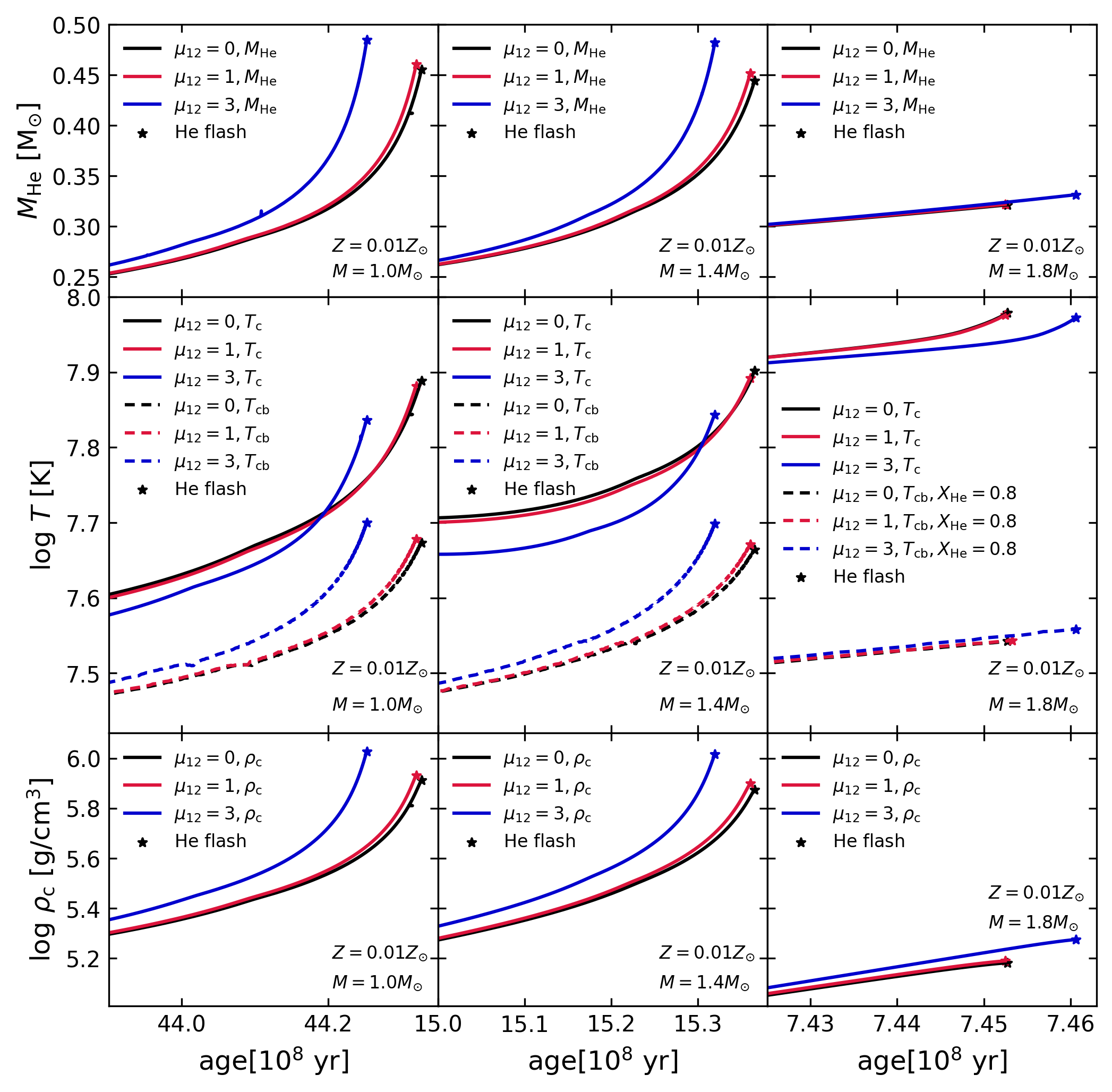} 
	\end{minipage}
	\caption{The structures and evolutions of models with different NMM values at $Z = 1 Z_{\odot}$, 0.01 $ Z_{\odot}$ and $M$ = 1.0, 1.4, 1.8 $M_{\odot}$ before the helium flash. The left figure shows the models with \( Z = 1\,\ Z_{\odot} \), while the right figure shows the model with \( Z = 0.01\,\ Z_{\odot} \). The x-axis in all panels represents the age in $10^8$ years. In the top panel, the y-axis shows the helium core mass. In the middle panel, the y-axis depicts both the central temperature and the core boundary temperature, presented on a logarithmic scale. The y-axis in the bottom panel represents the central density, also shown on a logarithmic scale.}
	\label{fig:1}
\end{figure*}
\begin{figure}[htbp]
	\centering
	\includegraphics[width=\linewidth]{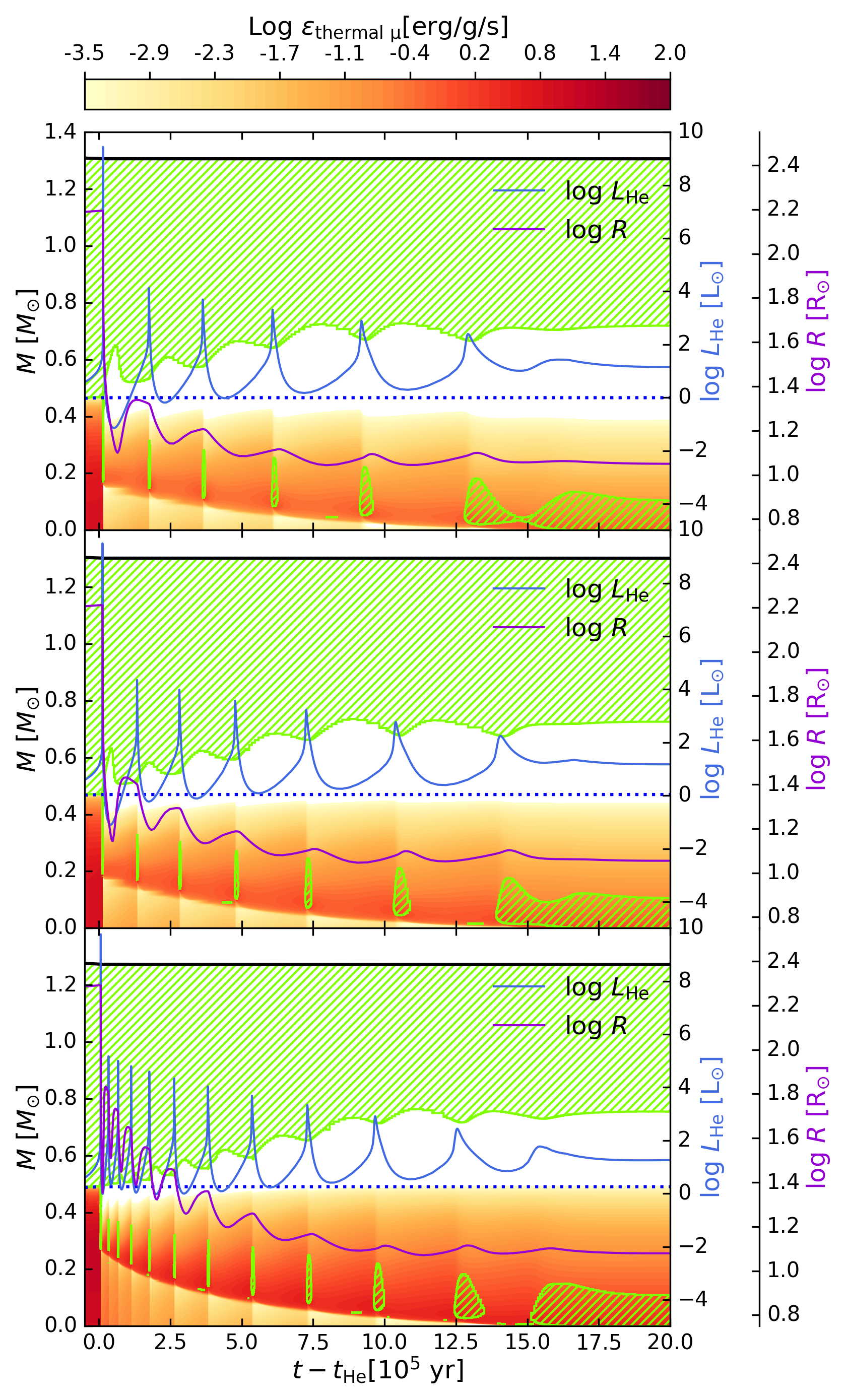}    
	\caption{The structures and evolutions of models with different NMM values at $Z = 1 Z_{\odot}$and $M$ = 1.4 $M_{\odot}$ during the helium flash phase. The NMM values for the models in the top, middle, and bottom panels are set to ${\mu_{12}}=0.0$, ${\mu_{12}}=1.0$, ${\mu_{12}}=3.0$, respectively. $t_{He}$ represents the time of the helium flash occurs. The green shaded area indicates the convection zone, the orange shaded area shows the neutrino production rate in logarithmic coordinates, the dotted line marks the helium core boundary, and the blue and purple lines represent the helium luminosity and stellar radius, respectively, each corresponding to their own separate y-axes on the right.}
	\label{fig:2}
\end{figure} 
\section{Result}
We evolve models with masses ranging from 1 to 2 $M_{\odot}$ and discuss the impact of different NMM models on the evolution of red giant stars before the helium flash, during the helium flash phase, and after the helium flash in the red clump phase. Considering the uncertainties in stellar evolution theory based on current observational constraints, our upper limit for NMM will be slightly higher than the current observational upper limit. Therefore, we adopt NMM values of ${\mu_{12}}= 0, 1, 3$, where ${\mu_{12}}= =\mu_\nu /\left(10^{-12} \mu_{\mathrm{B}}\right)$. Additionally, we explore the influence of varying gravity wave mixing parameters on the surface lithium abundance.

\subsection{The Influences of NMM on the Stellar Structure Befroe Helium Flash}
\label{section:3.1}
We find that models with identical metallicity and mass, but with varying NMM values, exhibit significant evolutionary differences in the red giant phase before the helium flash. Figure \ref{fig:1} shows the structures and evolutions of models with different NMM values at $Z = 1 Z_{\odot}$, 0.01 $Z_{\odot}$ and $M$ = 1.0, 1.4, 1.8 $M_{\odot}$ before the helium flash.
For the model with $Z = 1 Z_{\odot}$, as shown in the upper panel of the left figure in Figure \ref{fig:1}, the helium core mass ($M_{\mathrm{He}}$) increases at the helium flash as the model adopts a larger NMM. This finding is consistent with the results reported by \cite{Arceo2015APh,Giunti2016AnP,Mori2021}. Following \cite{Iben1993PASP}, We define the helium core boundary where the helium abundance reduces to 0.5 ($X_{\mathrm{He}}=0.5$).
Notably, contrary to \cite{Arceo2015APh} and \cite{Giunti2016AnP} conclusion that increasing NMM would delay the helium flash, Figure \ref{fig:1} indicates that the helium flash occurs earlier as the NMM increases. \cite{Mori2021} also reached the same result but did not provide a detailed explanation.

Firstly, the increase in $M_{\mathrm{He}}$ at the helium flash is primarily due to the additional energy loss by NMM. As illustrated in the middle panel of Figure \ref{fig:1}, where the central temperature ($T_{\mathrm{c}}$) of the star before the helium flash decreases as the model adopts a larger NMM. 
This is due to the degenerate helium core in the temperature range of ${10^{7.4}}-{10^8} K$ producing neutrinos and losing energy through plasma processes($\gamma^* \longrightarrow \nu+\bar{\nu}$), and the additional neutrino energy loss reduces $T_{\mathrm{c}}$. The plasma process enhances the reaction rate with temperature($P_{\mathrm{e}} \propto \rho^{\frac{5}{3}}$), so the model with a higher $T_{\mathrm{c}}$ of temperature($\varepsilon_{\mathrm{plas}}\propto{T^3}$), models with higher central temperatures, such as the $M = 1.8 M_{\odot}$ and different NMM values has a larger central temperature difference($\Delta T_{\mathrm{c}}$). The decrease in $T_{\mathrm{c}}$ results in a decrease in central gas pressure, causing the electron degeneracy pressure($P_{\mathrm{e}}$) in the helium core to rise to maintain gravitational balance, and since electron degeneracy pressure is proportional to density($P_{\mathrm{e}} \propto \rho^{\frac{5}{3}}$), this leads to an increase in central density ($\rho_{\mathrm{c}}$).
As shown in the bottom panel of Figure \ref{fig:1}, the $\rho_{\mathrm{c}}$ before the helium flash of the star increases as the model adopts a larger NMM, and similarly to $T_{\mathrm{c}}$, the difference is more pronounced for higher mass models.
Moreover, based on the polytropic model, helium core mass is proportional to central density($M_{\mathrm{He}} \propto \rho^{\frac{1}{2}}$), and a higher $\rho_{\mathrm{c}}$ implies that the star requires a helium core with higher degeneracy and larger $M_{\mathrm{He}}$ to trigger the helium flash.
Furthermore, a higher $\rho_{\mathrm{c}}$ indicates more intense contraction of the helium core, and the faster contracting helium core releases more gravitational potential energy, heating the hydrogen burning shell at the helium core boundary and causing the hydrogen shell layer temperature to rise. As shown by the dotted line in the middle panel of Figure 1, we observe that, contrary to $T_{\mathrm{c}}$, the temperature at the helium core boundary also increases with increasing NMM.
Similarly, due to the sensitivity of nuclear reaction rates($\varepsilon_{\mathrm{H}}$) to temperature($\varepsilon_{\mathrm{P-P}}\propto{T^5}$ ; $\varepsilon_{\mathrm{CNO}}\propto{T^{15}}$), the $\varepsilon_{\mathrm{H}}$ at the helium core boundary also increases with increasing NMM. Therefore, the increase in $\varepsilon_{\mathrm{H}}$ causes $M_{\mathrm{He}}$ to grow faster to the critical mass that triggers the helium flash, resulting in our models evolving faster during the red giant phase and experiencing the helium flash earlier with increasing NMM.  

\begin{figure}[htbp]
	\centering
	\includegraphics[width=\linewidth]{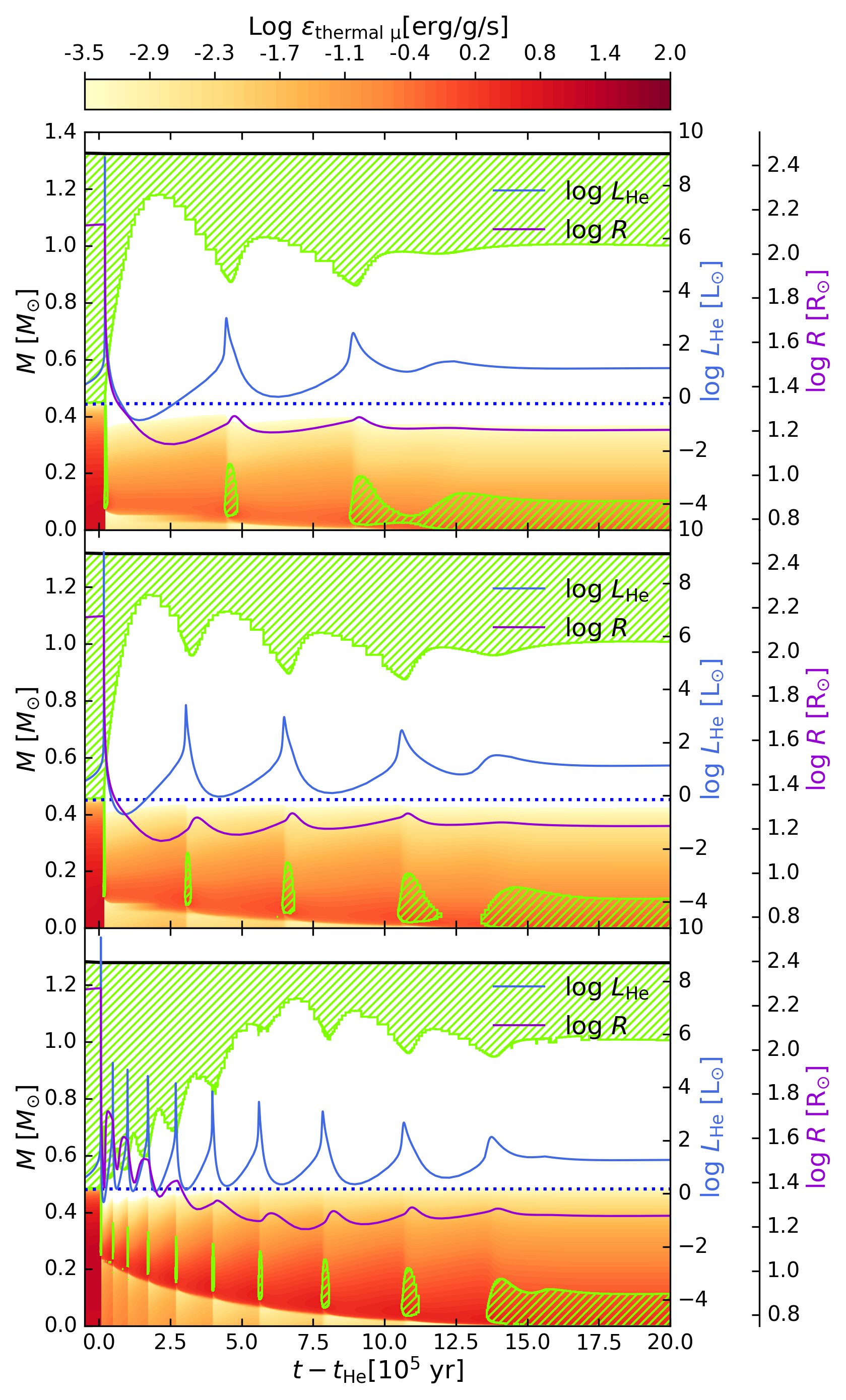}    
	\caption{Same as Fig. \ref{fig:2}, but for the models with $Z = 0.01 Z_{\odot}$.}
	\label{fig:3}
\end{figure}
For the model with an extremely low metal abundance of  $Z = 0.01 Z_{\odot}$, as shown in the right panel of Figure \ref{fig:1}, the evolution of the $1.0 M_{\odot}$ and $1.4  M_{\odot}$ models is similar to that of the  $Z = 1 Z_{\odot}$ model, which aligns with the property that the degenerate core depends only on its helium core mass. However, for model with $Z = 0.01 Z_{\odot}$ and $M = 1.8  M_{\odot}$, the star does not undergo a typical helium flash. This is because the central temperature is too high, causing helium ignition to occur before the central helium core is fully degenerate. As illustrated in the right panel of Figure \ref{fig:1}, at the time of the helium flash, the central temperature has already reached the helium ignition temperature($\sim{10^8}K$), and the central density is much lower than in other models. The mass of the helium core during the flash is only $0.3  M_{\odot}$, further indicating that the helium flash occurs before the core is fully degenerate. Due to the incomplete degeneracy of the helium core, its boundary is also located more internally compared to a fully degenerate helium core.
It can be observed that the region with a helium abundance of 0.8 within the temperature range for hydrogen burning, and there is a significant nuclear reaction rate, as shown in the right panel of Figure  \ref{fig:1}.

Overall, as the model adopts a larger NMM, the critical $M_{\mathrm{He}}$ required for the helium flash increases due to the additional energy loss through neutrinos in the helium core center. Conversely, the additional energy loss accelerates the contraction of the helium core, releasing gravitational energy that heats the hydrogen shell and increases $\varepsilon_{\mathrm{H}}$, thereby causing the helium core to reach the critical mass more quickly, resulting in an earlier helium flash.
\begin{figure*}
	\centering
	\includegraphics[width=\linewidth]{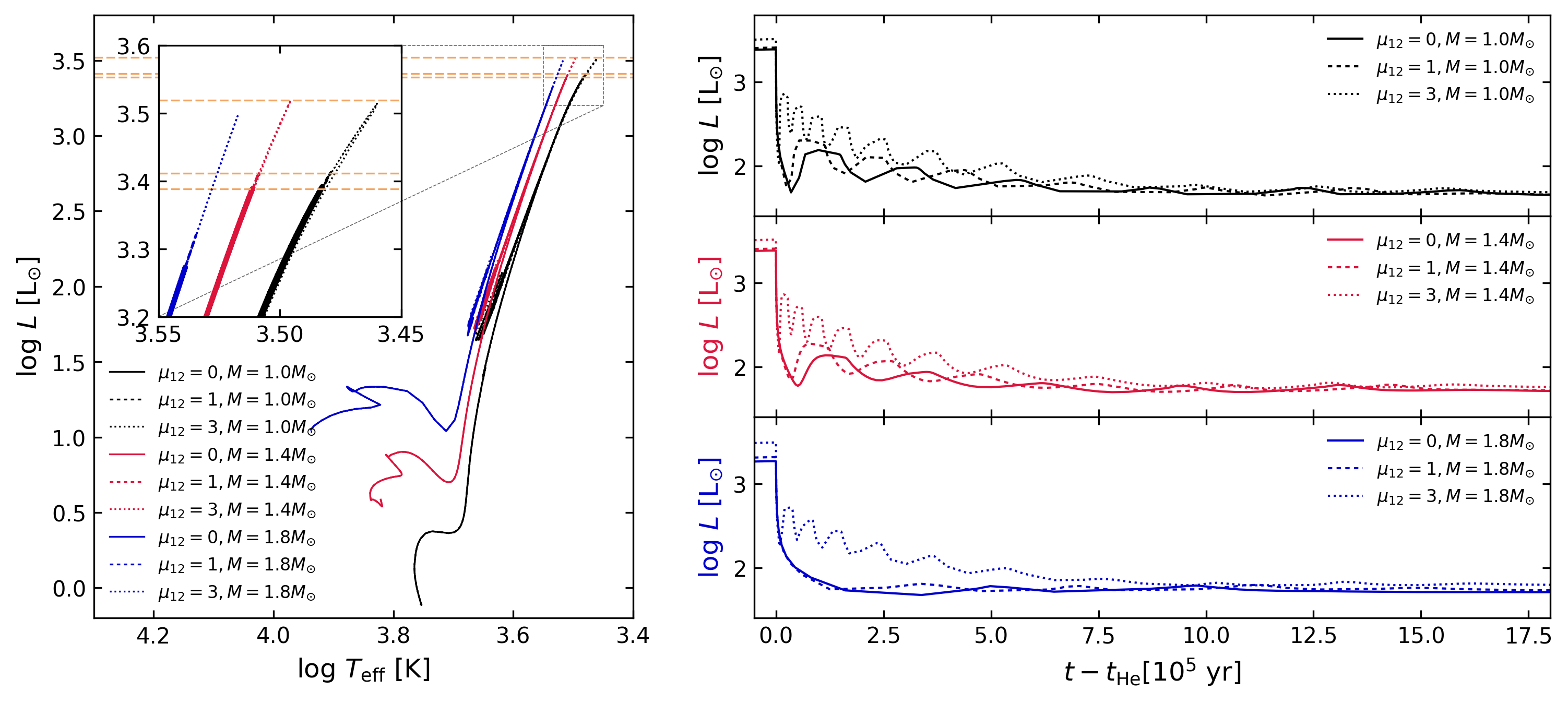}    
	\caption{Hertzsprung-Russell (HR) diagram and time-luminosity diagram for  $Z = 1 Z_{\odot}$ stars with different masses and NMM values. $t_{He}$ represents the time of the helium flash occurs. Black, red, and blue lines represent models with $M = 1.0, 1.4, 1.8$ $M_{\odot}$, respectively. Solid, dashed, and dotted lines represent models with ${\mu_{12}}=0.0, 1.0, 3.0$, respectively.}
	\label{fig:4}
\end{figure*}

\subsection{The Influences of NMM on Helium Flash}
\label{section:3.2}
Current research indicates that the helium flash process comprises the first helium flash (with ${L_{\mathrm{He}}} > {10^9} {L_{\mathrm{\odot}}}$ ) and a series of sub-flash\cite{Thomas1967ZA,Bildsten2012ApJ,Capelo2023ApJ}. Figure \ref{fig:2} illustrates the structural evolution of $1.4\, M_{\odot}$ stars with different NMM values during the helium flash. As depicted in Figure \ref{fig:2}, the first helium flash is ignited off-center, and each sub-flash process generates a convection zone in the burning region. This occurs because the maximum temperature in the helium core, due to neutrino energy loss at the core center, deviates from the center, resulting in off-center ignition. The substantial energy produced by the helium flash also causes the temperature gradient in the burning region to exceed the adiabatic gradient, leading to convection\cite{Serenelli2005A&A,Bildsten2012ApJ}.
As shown by the purple line and the green shaded area on the surface in Figure \ref{fig:2}, after each sub-flash, the stellar radius rapidly decreases, and the convective envelope also recedes towards the surface. This is because the degenerate helium core rapidly expands, pushing hydrogen burning outward and cooling down, thereby causing $\varepsilon_{\mathrm{H}}$ to rapidly decrease. Since the star's surface luminosity is provided by hydrogen burning at this time, the star's luminosity and radius also rapidly decrease, and the convective envelope recedes towards the surface as the radius decreases.

As shown by the blue line and the green shaded area in the helium core in Figure \ref{fig:2}, as the model adopts a larger NMM, the peak luminosity of the first helium flash is higher and more off-center, and the peak luminosity of subsequent sub-flash is higher with an increased number of sub-flash, and the time interval between sub-flash decreases. Due to the additional energy loss brought by the model adopting a larger NMM, the maximum temperature in the helium core deviates more from the center. Since a higher NMM produces a lager $M_{\mathrm{He}}$(see Figure \ref{fig:1}), the helium core $\rho_{\mathrm{c}}$ and degeneracy are higher, which results in higher peak luminosities for each helium flash.
Due to the additional neutrino energy loss, the region where each helium flash releases degeneracy and ignites becomes smaller, requiring more sub-flash to extend the burning region to the center of the helium core. As shown by the orange shaded area in Figure \ref{fig:2}, the model with a larger NMM value has a larger neutrino energy loss in the core burning region. Moreover, since the increased burning area decreases each time, the thermal equilibrium timescale of the new burning area also decreases accordingly, resulting in a shorter time interval between sub-flashes, and the overall timescale of the helium flash process remains essentially unchanged.

For the $Z = 0.01 Z_{\odot}$ model, as shown in Figure \ref{fig:3}, the evolution of the helium flash process is consistent with the results of the $Z = 1 Z_{\odot}$ model. Due to the extremely low metallicity, the helium abundance in the helium core of the model is higher, resulting in higher degeneracy pressure at the same $\rho$, thus the critical $M_{\mathrm{He}}$ is lower. Consequently, the peak luminosity of the helium flash is lower than that of the $Z = 1 Z_{\odot}$ model, and the number of sub-flashes is also fewer.
\begin{figure*}[htbp]
	\centering
	\includegraphics[width=\linewidth]{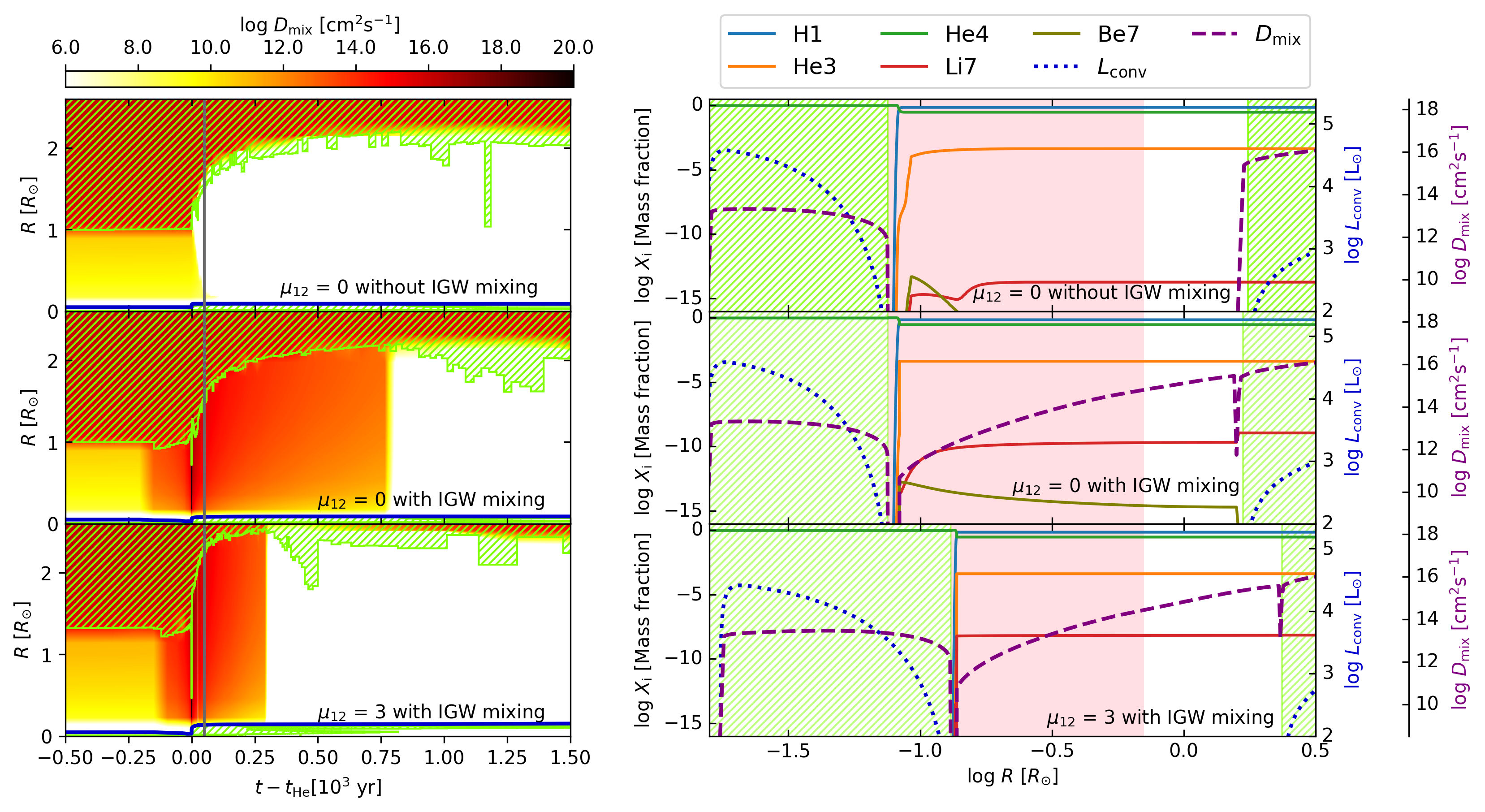}    
	\caption{the structural evolution of $Z = 1 Z_{\odot}$, $M = 1.0 M_{\odot}$ star models with different NMM values and with and without IGW mixing during the first helium flash, along with the internal structure at 50 years after the helium flash. In the left panel, the blue line represents the hydrogen-burning zone at the helium core boundary, and the vertical gray line marks the 50-year time point selected for the right panel. The green shaded area represents the convection zone, the orange shaded area in the left panel indicates the mixing coefficient, and the pink shaded area in the right panel indicates the Li-burning zone.}
	\label{fig:5}
\end{figure*}

\subsection{The Influences of NMM on Tip of the Red Giant Branch} 

Due to the consistent luminosity of the tip of the red giant branch, it can serve as a standard reference for distance measurement\cite{Bellazzini2001ApJ,Arceo2015APh,Jang2017ApJ,Freedman2020ApJ,Anand2022ApJ}. The helium flash corresponds to the evolutionary stage from the TRGB to the RC, and it is evident that the value of NMM will influence this stage. Figure \ref{fig:4} shows the Hertzsprung Russell (HR) diagram and the luminosity evolution over time for models with $Z = 1 Z_{\odot}$, different masses, and different NMM values. Based on the HR diagram in Figure \ref{fig:4}, it is shown that the luminosity of the TRGB increases with increasing NMM, and the model with $M = 1.0 M_{\odot}$ and ${\mu_{12}}=3.0$ shows a luminosity increase of $\sim 35\%$ ( from $ \sim 2450 L_{\odot}$ to$ 3320 L_{\odot}$ ) compared to its ${\mu_{12}}=0.0$ counterpart, in agreement with the \cite{Arceo2015APh}.
The luminosity of the TRGB is dependent on the $\varepsilon_{\mathrm{H}}$ in the hydrogen shell, as discussed in Sec \ref{section:3.1}, where the $\varepsilon_{\mathrm{H}}$ in the hydrogen shell is proportional to the temperature, and the temperature of the hydrogen shell is proportional to the $M_{\mathrm{He}}$. Therefore, models with larger NMM have larger critical $M_{\mathrm{He}}$ (see Figure \ref{fig:1}), which results in an increase in the luminosity of the TRGB. The right panel of Figure \ref{fig:4} shows that models with larger NMM exhibit a significantly smaller decrease in luminosity after the first helium flash, and the number and amplitude of subsequent luminosity changes increase with the number and amplitude of sub-flashes.
As discussed in Sec \ref{section:3.2}, larger NMM causes the first helium flash to be more off-center, resulting in a smaller region where the helium core releases degeneracy and less radius expansion, leading to a smaller cooling amplitude in the hydrogen shell. Due to the increase in the number and intensity of subsequent sub-flashes, the number and intensity of helium core degeneracy releases also increase, resulting in an increase in the number and amplitude of stellar luminosity changes.
\begin{figure}[htbp]
	\centering
	\includegraphics[width=\linewidth]{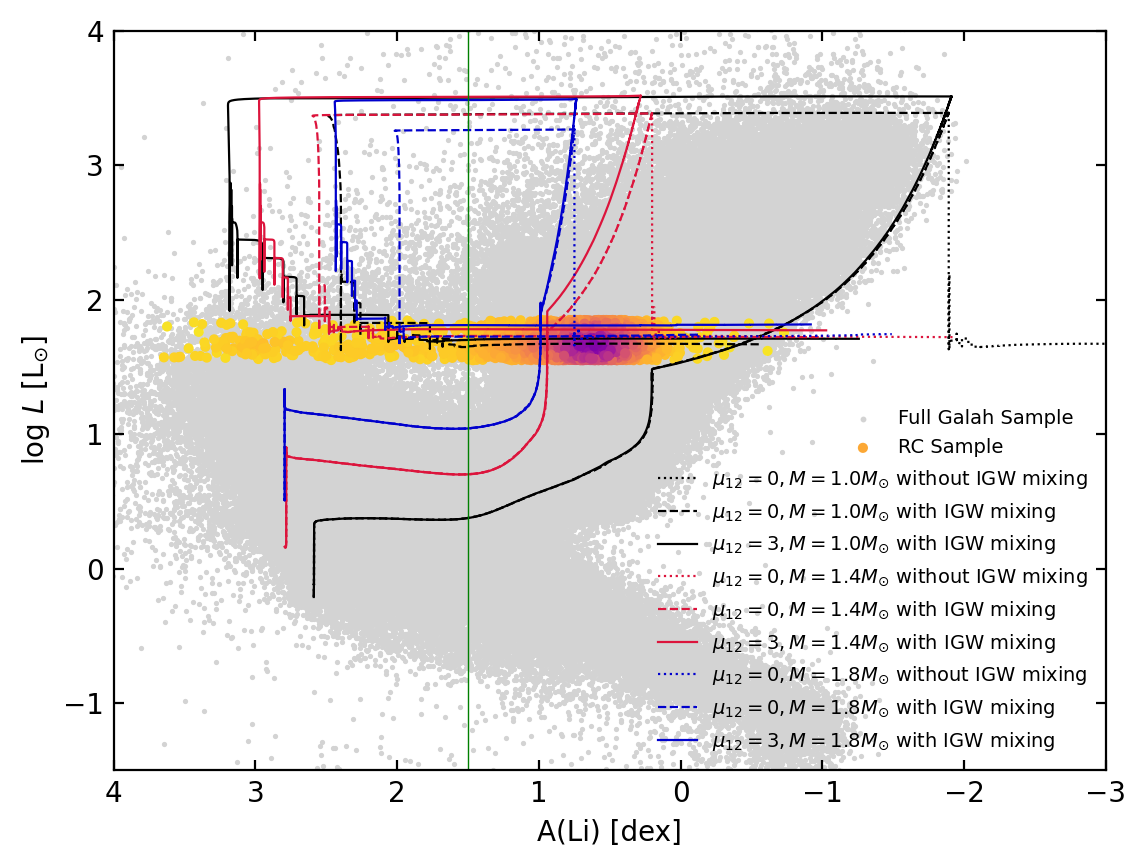}    
	\caption{The evolution of luminosity and surface A(Li) for $Z = 1 Z_{\odot}$ models with varying parameters. Color line indicates the model mass (black for $M = 1.0 M_{\odot}$, red for $M = 1.4 M_{\odot}$, blue for $M = 1.8 M_{\odot}$). Line style indicates the model parameters (dotted line for ${\mu_{12}}=0.0$ without IGW mixing, dashed line for ${\mu_{12}}=0.0$ with IGW mixing, solid line for ${\mu_{12}}=3.0$ with IGW mixing). The green line demarcates the boundary between Li-rich RCs and normal RCs. Colored dots show the RC samples from \citet{Kumar2020NatAs}, where the color scale indicates sample density (purple representing the highest density). Gray dots show the full samples complete from GALAH DR4 \cite{Buder2024arXiv}.}
	\label{fig:6}
\end{figure}

\subsection{The impact of IGW mixing on Li abundance}
Some studies propose that the Li in Li-rich RCs originates from the CF mechanism (${ }^3 \mathrm{He}(\alpha, \gamma)^7 \mathrm{Be}(e, v)^7 \mathrm{Li}$) in the hydrogen-burning zone\cite{Cameron1971ApJ}, but the lack of mixing in the radiative zone outside the burning zone is the most challenging issue for this theory\cite{Denissenkov2024MNRAS}. IGWs excited by random turbulent motion at the boundary of the convection zone can induce mixing in the radiative zone, a phenomenon supported by various theories and 3D simulations\cite{Edelmann2019ApJ,Herwig2023MNRAS,Thompson2024MNRAS}. The intense convection in the core triggered by the helium flash will generate strong IGWs, which will induce strong mixing in the radiative zone.

To investigate the impact of IGW mixing, we use Figure \ref{fig:5} to analyze the structural evolution of 
$Z = 1 Z_{\odot}$, $M = 1.0 M_{\odot}$ star models with different NMM values and with and without GW mixing during the first helium flash (left panel), both elemental abundance, convective luminosity, and $D_{\text {mix}}$ as a function of radius at 50 years after the helium flash (right panel). As shown in the left panel of Figure \ref{fig:5}, for models without IGW mixing, the radiative zone outside the core has a thermal diffusion mixing coefficient $K \sim 10^{8} \mathrm{~cm}^2 \mathrm{~s}^{-1}$ before the helium flash, but $K$ rapidly decreases to below $10^{6} \mathrm{~cm}^2 \mathrm{~s}^{-1}$ after the helium flash.
It is precisely because of the $K \sim 10^{8} \mathrm{~cm}^2 \mathrm{~s}^{-1}$ in the radiative zone before the TRGB that Li from the surface convection zone diffuses into the Li-burning zone ($\gtrsim 2.5 \times 10^6 \mathrm{~K}$), resulting in the Li depletion phenomenon in red giants. After the helium flash, due to the expansion of the helium core, the radiative zone outside the core rapidly cools, causing $K$ to decrease rapidly ($K \propto T^{3}$). Consequently, after the helium flash, there is no change in the elemental abundance of the envelope, as shown in the right-top panel of Figure \ref{fig:5}.

As shown in the left-middle panel of Figure \ref{fig:5}, for the ${\mu_{12}}=0.0$ model with IGW mixing, the IGWs excited after the helium flash induce intense mixing in the radiative zone for a duration of $\sim 80$ years. 
Press performed an energy budget analysis showing that mixing in a given region requires sufficient energy to overcome the local gravitational potential barrier, with mixing efficiency evolving on thermal diffusion timescales\cite{Press1981ApJ}. This energy requirement corresponds to a characteristic power of $\sim N^2 K$. Within the mixing region, this power is estimated to be of order $\sim 10^7 \mathrm{erg} \, \mathrm{s}^{-1} \mathrm{~g}^{-1}$ \cite{Schwab2020}. Given the typical IGW mixing region mass of $\sim 10^{30} \mathrm{g}$, the critical luminosity for IGW mixing initiation is approximately $L_{\text {crit }} \sim 10^4 L_{\odot}$.
Within As shown in the right panel of Figure \ref{fig:5}, the internal convective luminosity induced by the helium flash exceeds the critical luminosity for IGW mixing, resulting in a high mixing coefficient $D_{\text {mix}}$ in the radiative zone, and the Li produced by the Cameron-Fowler (CF) mechanism in the hydrogen-burning zone is thus mixed into the surface convection zone. 
For the ${\mu_{12}}=3.0$ model with IGW mixing, the IGW mixing excited by the helium flash lasted for a shorter duration of ~30 years. As discussed in Sec \ref{section:3.2}, the additional energy loss by NMM causes the first helium flash to be more off-center, resulting in a smaller region where degeneracy is released.

Notably, as shown in the right panel of Figure \ref{fig:5}, the ${\mu_{12}}=3.0$ model with IGW mixing has almost the same internal convective luminosity and radiative zone $D_{\text {mix}}$ as the ${\mu_{12}}=0.0$ model with IGW mixing, but the Li abundance in the surface convection zone is higher, indicating that more Li in the radiative zone is mixed into the surface convection zone. As discussed in Sec \ref{section:3.2}, a larger NMM brings a higher peak luminosity of the helium flash, and a higher peak luminosity of the helium flash causes the core to expand faster initially, and the surface convection zone to retreat towards the surface faster. A faster retreating surface convection zone results in a narrower overshoot region at the boundary of the surface convection zone.
As shown in the right panel of Figure \ref{fig:5}, both the ${\mu_{12}}=0.0$ with IGW mixing model and the ${\mu_{12}}=3.0$ with IGW mixing model have a rapidly  decreasing concave region of $D_{\text {mix}}$ at the boundary of the surface convection zone, which is the overshoot region of both models. Since theoretically only the propagation of IGWs in a stable radiative zone is considered, overshoot region becomes a "bottleneck" region with low $D_{\text {mix}}$. Since there is a Li-burning zone in the radiative zone (shown by the pink shaded area in the right panel of Figure \ref{fig:5}), a smaller overshoot region brought by a larger NMM means that Li mixes the surface convection zone faster, stays in the radiative zone for a shorter time, and is consumed less by burning, so the ${\mu_{12}}=3.0$ model with IGW mixing has a higher surface Li abundance. The jump in Li abundance in the overshoot region of the ${\mu_{12}}=0.0$ with IGW mixing model also confirms this physical process.

Figure \ref{fig:6} shows the evolution of A(Li)-luminosity for $Z = 1 Z_{\odot}$ models with different parameters. 
As shown by the dotted line, the standard model without IGW mixing exhibits no change in A(Li) after the helium flash, and A(Li) gradually decreases as it enters the RC phase. This poses a challenge for the standard model, as low-mass stars undergo the first dredge-up when they evolve into the red giant phase, during which the surface convection zone rapidly extends into the stellar interior, bringing surface Li into the interior, leading to a rapid decrease in A(Li). Notably, all models show lithium depletion (A(Li)) starting at the RGB bump luminosity ($1.5 - 2.0 L_{\odot}$) until the TRGB, caused by thermohaline mixing activated by molecular weight inversion\cite{Angelou2012ApJ,Lattanzio2015MNRAS}. The standard model cannot explain either Li-rich RCs or even normal RCs with such low A(Li) values. As shown by the dashed line in Figure \ref{fig:5}, the standard model with IGW mixing demonstrates a significant increase in A(Li) after the helium flash. Since the strength of IGW mixing is related to the intensity of the helium flash, the lower the stellar mass, the more pronounced the increase in A(Li). For the $M = 1.0 M_{\odot}$ model, the helium flash increases A(Li) from -1.9 dex to 2.4 dex. The standard model with IGW mixing can account for most Li-rich RCs, but some super Li-rich RCs with A(Li) greater than 3.2 dex in Figure \ref{fig:6} remain challenging to explain. As discussed in the previous paragraph, using a larger NMM in the model can narrow the low-$D_{\text {mix}}$ bottleneck region caused by overshoot, thereby enhancing the enrichment effect of IGW mixing on surface Li. As shown by the solid line in Figure \ref{fig:6}, the ${\mu_{12}}=3.0$ model with IGW mixing demonstrates a significant increase in A(Li) after the helium flash. For the $M = 1.0 M_{\odot}$, ${\mu_{12}}=3.0$ model with IGW mixing, the helium flash increases A(Li) from -1.9 dex to 3.2 dex, encompassing the super Li-rich RCs sample.

As shown in Figure \ref{fig:6}, the models with IGW mixing demonstrate that stars become Li-rich RCs after the helium flash, then gradually decrease in surface lithium abundance over time, eventually returning to normal RCs. This is caused by gravity sedimentation, where the gravitational acceleration term in the diffusion equations gives heavy elements an inward radial diffusion velocity. In our IGW models with different masses, stars only appear as Li-rich RCs for 2-15 Myr after the helium flash, while the entire red RCs lasts about 110-130 Myr. This timescale predicts that Li-rich RCs should only account for a few percent of all Li-rich RCs, consistent with both the $D_{\text {mix }}=10^{13} \mathrm{~cm}^2 \mathrm{~s}^{-1}$ model results from \cite{Gao2022A&A} and the observed fraction of Li-rich RCs\cite{Kumar2020NatAs}. Our models also agree with recent observations\cite{Mallick2025ApJ} that Li-rich RCs are generally younger. Specifically, RCs right after the helium flash tend to have higher A(Li), which then decrease with age due to gravity sedimentation.
Additionally, we find that the duration of the Li-rich phase increases with stellar mass. The typical Li-rich phases last 3, 7, and 15 Myr for $1.0 M_{\odot}$, $1.4 M_{\odot}$, and $1.8 M_{\odot}$ models respectively. This suggests that more massive RCs have a higher fraction of Li-rich RCs, matching observed mass distribution of normal and Li-rich RCs \cite{Zhou2022ApJ}.

However, in recent years, a small number of Li-rich samples with even higher A(Li) that we cannot explain have been discovered, such as TYC 429-2097-1, which has an A(Li) of 4.51 \cite{Yan2018NatAs}. This could be due to the conservative selection of our IGW mixing parameters, or these extreme samples might originate from binary star mergers.

\section{Conclusions}
In this paper, we first examined the effect of additional energy loss by NMM on the helium flash. We then introduced IGW mixing, analyzed its impact and its combined effect with NMM, and finally compared the Li-rich RC star samples. We found that as the model adopts a larger NMM, the critical $M_{\mathrm{He}}$ required for the helium flash increases, consistent with \cite{Arceo2015APh,Giunti2016AnP}. For the most typical $Z = 1 Z_{\odot}$,$M$ = 1.0 $M_{\odot}$, and $\mu_v = 3 \times 10^{-12} \mu_{\mathrm{B}}$ model, the helium core mass is $\sim 5\%$ ( from $ \sim 0.466 M_{\odot}$ to $ 0.490 M_{\odot}$ ) larger than that of the model without NMM. However, unlike their conclusion that NMM would delay the helium flash, our model indicates that the helium flash occurs earlier with increasing NMM values. 
This occurs because the additional energy loss accelerates the contraction of the helium core, releasing gravitational energy that heats the hydrogen shell and increases $\varepsilon_{\mathrm{H}}$. As a result, the helium core reaches the critical mass faster, advancing the helium flash.

The increase in NMM results in a higher and more off-center peak luminosity for the first helium flash, as well as higher peak luminosities for sub-flashes with an increased number of occurrences, and a shorter time interval between sub-flashes. Regarding the impact of the helium flash on stellar luminosity, first, the luminosity of the TRGB also increases with increasing NMM, consistent with the \cite{Arceo2015APh}. For the typical $Z = 1 Z_{\odot}$ , $M$ = 1.0 $M_{\odot}$, and $\mu_v = 3 \times 10^{-12} \mu_{\mathrm{B}}$ model, the luminosity is $\sim 35\%$ higher ( from $ \sim 2450 L_{\odot}$ to $ 3320 L_{\odot}$) compared to the model without NMM. Second, the increase in NMM significantly reduces the amplitude of luminosity attenuation after the first helium flash, and the number and amplitude of subsequent luminosity changes increase with the number and amplitude of sub-flashes.

The ongoing advancements in asteroseismology enable probe the internal physical parameters of stars\cite{Yan2021NatAs,Maben2023ApJ}. This implies that the effects of NMM on the helium flash could be observed, thereby enabling stronger constraints on NMM.

The IGW mixing generated by the helium flash can induce sufficient mixing in the radiative zone to transport the Li produced by the stellar CF mechanism to the surface, but the low-Dmix bottleneck region caused by overshoot diminishes this effect. The increase in NMM in the model narrows the overshoot bottleneck region, enabling Li to enter the surface convection zone more quickly, thereby enhancing the enrichment effect of IGW mixing on surface Li. When comparing the Li-rich RC star samples, it is evident that models with only IGW mixing can generate Li-rich RC star samples but fail to produce super Li-rich RCs. 
For models incorporating both NMM and IGW mixing, the duration of IGW mixing is reduced. However, due to the narrowing of the overshoot bottleneck region, Li is mixed into the surface convection zone more quickly, reducing the time Li spends in the radiative zone and decreasing the amount of Li consumed by burning, thereby effectively producing super Li-rich RCs.

Given the relatively high precision of surface elemental abundance measurements in stars, NMM's influence on the helium flash, and its enhancement of IGW mixing effects, future observations of RC stars with anomalous surface element abundances may offer a new approach to constraining NMM.

\section*{Acknowledgements}
This work received the generous support of the National Natural Science Foundation of China under grants Nos. 12163005, U2031204, 12373038, and 12288102, the science research grants from the China Manned Space Project with No. CMSCSST-2021-A10, and the Natural Science Foundation of Xinjiang Nos. 2022D01D85 and 2022TSYCLJ0006.

\section{Data available}
The files to reproduce our work are publicly available at: \url{https://doi.org/10.5281/zenodo.15172801}

\bibliography{paper}

\end{document}